\documentclass[
reprint,
longbibliography,
 amsmath,amssymb,
 aps,
floatfix,
]{revtex4-2}

\usepackage{blindtext}

\usepackage{adjustbox}
\usepackage{subfigure}
\usepackage{amssymb}
\usepackage{upgreek}
\usepackage{verbatim}
\usepackage[separate-uncertainty=true]{siunitx}

\usepackage{graphicx}
\usepackage{dcolumn}
\usepackage{bm}
\usepackage{xcolor}

\begin{document}

\preprint{APS}

\title{Dynamic convergent shock compression initiated by return current in high-intensity laser solid interactions}

\author{Long Yang\textsuperscript{1,2}}
\email{yanglong@hzdr.de}
\author {Martin Rehwald\textsuperscript{1}} 
\author {Thomas Kluge\textsuperscript{1}}
\author {Alejandro Laso \textsuperscript{1}}
\author{Toma Toncian\textsuperscript{1}}
\author {Karl Zeil\textsuperscript{1}}
\author {Ulrich Schramm\textsuperscript{1,2}}
\author{Thomas E Cowan\textsuperscript{1,2}}
\author{Lingen Huang\textsuperscript{1}}

\affiliation{(1) Helmholtz-Zentrum Dresden - Rossendorf, 01328 Dresden, Germany}
\affiliation{(2) Technische Universit\"at Dresden, 01062 Dresden, Germany}

\date{\today}

\begin{abstract}
We investigate the dynamics of convergent shock compression in the solid wire targets irradiated by an ultra-fast relativistic laser pulse. Our Particle-in-Cell (PIC) simulations and coupled hydrodynamic simulations reveal that the compression process is initiated by both magnetic pressure and surface ablation associated with a strong transient surface return current with the density in the order of \SI{e17}{A/m^2} and a lifetime of \SI{100}{fs}. The results show that the dominant compression mechanism is governed by the plasma $\beta$, i.e., the ratio of the thermal pressure to magnetic pressure. For small radii and low atomic number Z wire targets, the magnetic pressure is the dominant shock compression mechanism. 
As the target radius and atomic number Z increase, the surface ablation pressure is the main mechanism to generate convergent shocks based on the scaling law. Furthermore, the indirect experimental indication of the shocked hydrogen compression is provided by measuring the evolution of plasma expansion diameter via optical shadowgraphy. This work could offer a novel platform to generate extremely high pressures exceeding Gbar (100 TPa) to study high-pressure physics using femtosecond J-level laser pulses, offering an alternative to the nanosecond kJ laser pulse-driven and pulse power Z-pinch compression methods.
\end{abstract}

\maketitle

\section{Introduction}
Dynamic shock compression is a pivotal method to generate warm and hot dense matter, akin to the extreme conditions naturally observed in planetary interiors, supernovae, and astrophysical jets. In the laboratory, a variety of large-scale facilities are used to drive the dynamic compression including gas guns \cite{fowles1970gas}, pulse power systems\cite{deeney1998enhancement,spielman1995pbfa,huang2017radiation}, and nanosecond high-energy laser pulses\cite{moses2010advances,spaeth2016description}. Pulsed power techniques stand out among these due to their ability to produce intense short-duration currents, approximately 100 ns in length, with peak values reaching into tens of MegaAmperes \cite{meng2021visualizing,bailey2002x,rochau2008radiating}. When these currents are introduced to a target, the resultant $\textbf{J}\times \textbf{B}$ force can accelerate it to speeds of up to 40 km/s within a 100 ns timeframe, a rate significantly higher than that achieved by traditional gas-gun methods \cite{lemke2005magnetically,knudson2008shock,root2015shock}.
Pulsed power platforms are versatile. They are used for investigating the equation of state via shock generation \cite{knudson2013adiabatic,ozaki2016dynamic} and for creating X-rays through convergent compression mechanisms, aiding both X-ray source studies and fusion research \cite{deeney1997power,spielman1998tungsten,cuneo2001development,rochau2007high}.
However, the use of 100 ns current pulses demands intricate target designs to prevent magnetic diffusion within the target and to address issues related to magneto-hydrodynamic (MHD) instabilities \cite{davis2005magnetically,li2019experimental,zhang2019sustained}. Several strategies have been introduced to tackle these challenges, including the integration of an external axial magnetic field to stabilize such instabilities \cite{wright1976stability,delzanno2007resistive,delzanno2008effect} and optimizing the target size to achieve quasi-isentropic compression to prevent magnetic diffusion \cite{kraus2016dynamic,bradley2009diamond,knudson2012megaamps}. While nanosecond kilo-joule level high-energy lasers with high cost can reach TPa pressures on targets \cite{abu2022lawson,jiang2018experimental,kritcher2008ultrafast}, recent advancements in femtosecond high-intensity lasers offer the promise of generating short pulse currents with durations that are potentially much shorter
than the usual unstable MHD modes \cite{haines2011review}. The femtosecond high-intensity lasers could serve as an alternative way to initiate convergent shock which is less susceptive to hydrodynamic instabilities.


Here, we propose to investigate the utility of femtosecond relativistic laser as a novel approach for initiating cylindrically converging shocks. When a femtosecond relativistic laser is incident on a solid target, energetic hot electrons with kinetic energy exceeding MeV are generated through the Lorentz force\cite{stupakov2001ponderomotive}. The subsequent escape of hot electrons from the target leads to the formation of a surface return current\cite{PhysRevLett.117.035004,PhysRevLett.85.2945,beg2004return,hauer1983return,benjamin1979measurement} and a corresponding tens of kiloteslas magnetic field \cite{sandhu2002laser}. Previous research on the return current has largely concentrated the conventional z-pinch effect, which employs the substantial $\textbf{J}\times \textbf{B}$ force to continuously drive the target to high densities and temperatures 
\cite{PhysRevLett.117.035004,hauer1983return,ong2023ultra,beg2004high}. In contrast, our focus will be on utilizing the transient nature of the return current to initiate converging shocks by transient $\textbf{J}\times \textbf{B}$ force and surface joule heating. The return current
lasts within 100 fs timescales and  is constrained within the target’s skin depth
and hence, making the magnetic diffusion negligible, allowing the bulk of the target to maintain a relatively cold state. Also, MHD instabilities do not have sufficient time to develop within such an ultrashort period\cite{haines2011review}. The return current can rapidly increase the target surface temperature via joule heating, and the resultant transient pressure gradient can generate strong shocks.
plasma ablation can further compress the target. 
Consequently, the converging shock benefits from the transient current duration, 
reducing the need for complex designs 
to suppress the MHD instability and to achieve symmetrically converging shocks.

In this study, we investigate the dynamic shock compression initiated by surface return current using a multi-timescale methodology. To elucidate the mechanisms involved in this process, we employ a comprehensive approach that combines Particle-in-Cell (PIC) simulations, and hydrodynamic simulations. Our focus is on examining the compression effect from hydrogen to high Z metallic wires. The PIC method allows us to understand the mechanism of compression initiation while the hydrodynamic simulations offer a robust approach to model the effect in longer times. Hydrogen, with its single electron and proton, is ideally suited for simulation with the PIC method. Its low density enables clear observation of the compression effect from the PIC method. We demonstrate that the dynamic shock compression initiated by the return current is a consequence of the interplay between magnetic compression and surface ablation. In radii of a few micrometers, the shock compression is dominated by magnetic compression. In a \SI{2.5}{\um} hydrogen jets target, the surface plasma is heated to 300 eV in 100 fs by a \SI{e17}{A/m^2} surface return current. The peak thermal and magnetic pressures correspond to 60 TPa and 20 TPa, respectively, at the peak of the surface return current at 240 fs. The inward converging shock velocity can reach 800 km/s. The converged shock collapses to the target center at about 10 ps. The hydrogen density is compressed by a factor of 29 with 220 eV temperature and 100 TPa pressure. A predicted rebounded shockwave in the solid hydrogen jets is indirectly indicated experimentally via shadowgraphy images \cite{Martin2022}. We scale the return current dynamic compression to materials with high Z and larger radii. It is found that surface ablation is the main mechanism to initiate a convergent shock in large radii and high Z wire targets. In a \SI{10}{\um} copper wire, the thermal pressure is 38 times of magnetic pressure initially. The pressure at the target center when imploded is 125 TPa. The PIC simulation with  \SI{10}{\um} diameter copper wire is applied as an example to verify the return current scaling law.
These findings establish the return current compression as a promising novel platform for future high-pressure physics research.\\

The structure of this paper is as follows: Section II discusses the generation of return current and its subsequent impact on the target, accompanied by a multi-timescale analysis to examine the mechanism underlying the dynamic compression initiated by the return current. In Section III, experimental data of the evolution of plasma shadowgraphy diameter are provided, as captured by shadowgraphy imaging techniques. The results might hint towards a combination of surface ablation and the $\textbf{J} \times \textbf{B}$ force originating from a laser-initiated return current. In Section IV, we have a conclusion to summarize this study.

\section{Dynamic shock compression initiated by the surface return current}
The surface return current has two primary effects on a wire target: $\textbf{J}\times \textbf{B}$ force and joule heating.  These two effects will be discussed in Subsections A and B, respectively. In Subsection C, the converging shock propagation in \SI{100}{\ps} time scale is presented. In Subsection D, a return current scaling law is introduced to evaluate the convergent shock compression mechanisms under conditions where the laser is incident to high atomic numbers (Z) and large-radius targets.

\subsection{The z-pinch effect induced by the strong transient magnetic field}
\begin{figure*}[htb] 
\centering {
\includegraphics[width=0.9\textwidth]{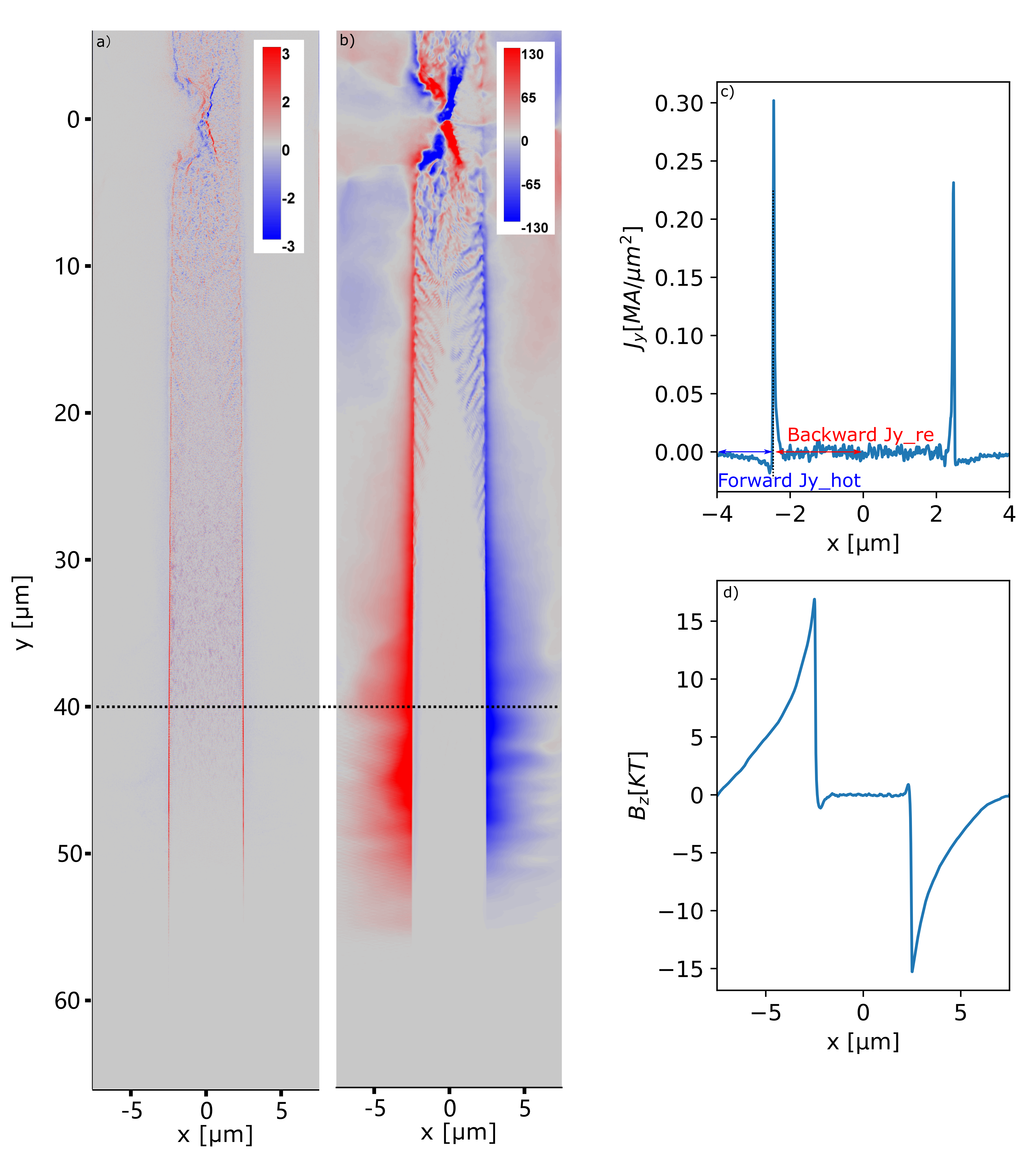}} 
\caption{PICLS 2D simulation of return current formation in a solid hydrogen wire at 160 fs. a) The 2D distribution of current density at y direction, the unit of the current density is $en_cc$ (\SI{0.08352}{MA/\um^2}). b) The 2D distribution of the magnetic field in the z-direction, the unit of the magnetic field is 100 Tesla. c) The lineout of return current density at y=\SI{40}{\um} in figure (a). d) The lineout of $B_z$ at y=\SI{40}{\um} in figure (b).}
\label{Fig.picls_jy10}
\end{figure*}
In this subsection, we analyze the magnetic compression initialized by the surface return current.
We adopt the concept of plasma $\beta$, which serves as a criterion to determine whether the plasma compression is dominated by a magnetic field or thermal pressure. The plasma $\beta$ is defined as $\beta = P_T/P_B$, where $P_T = n_e k_B T_e$ represents the surface ablation plasma pressure and $P_B = B_{\theta}^2/2\mu_0$ denotes the surface magnetic pressure. The initial step involves the deviation of the surface return current from the PIC simulations in order to compute the plasma $\beta$.\\
Particle-in-Cell (PIC) simulations are a superior tool for investigating short-pulse laser-solid interactions. In this study, we apply PIC simulations to explore the surface return current dynamics in wire targets. The comprehensive setup details of PIC simulations are elucidated in Appendix A. Figure \ref{Fig.picls_jy10} (a) and (b) show the distributions of current density and magnetic field,  respectively, in the PICLS simulations at 240 fs, illustrating the propagation of a current pulse on the surface of the plasma column. The spatial extent of this current pulse is approximately \SI{30}{\um}. As the current pulse propagates approximately with the speed of light\cite{PhysRevLett.102.194801}, the temporal duration of the return current pulse is estimated to be around 100 fs. Figure \ref{Fig.picls_jy10} (c) and (d) demonstrate the lineout of the current density and magnetic field at y = \SI{40}{\um} (\SI{40}{\um} from the laser spot), where the current density and magnetic field on the plasma reach their peak intensity. The average peak current density is \SI{0.28}{MA\um^{-2}}, and the average peak magnetic field is 19 kT. 
Similarly, the results of the 3D PIC simulations are exhibited in Figure \ref{Fig.3DPIConGPU} in Appendix A.  
In this scenario, the peak current density is considered to be at a distance of \SI{5}{\um} from the laser spot region. The corresponding current density and magnetic field are \SI{0.9}{MA\um^{-2}} and 46 kT, as shown in Figure \ref{Fig.3DPIConGPU} (e) and (f).\\

The surface current has several components: the escaping hot electron current ($I_{es}$) which is a negligible proportion of total laser-accelerated electrons \cite{huang2022dynamics}, the forward-propagating laser-accelerated electron current ($I_{h}$) outside the plasma column, and the backward surface return current ($I_{re}$) in the skin depth of the target that neutralizes the forward laser-accelerated electron current and escaping electron current. Thus, we have the equation $I_{re} \simeq I_{h}$. Here, the forward current is the current moving along the wire surface away from the interaction area, and the return current is moving toward the interaction area. The forward electron current and backward surface return current are in different layers along the radial direction. The interface of these two layers is around the plasma surface (r $\sim$ \SI{2.5}{\um} in 2D PIC simulations).
The current is calculated by integrating the current density
over the radius radially, as depicted in Equation (2).\\

\begin{figure}[htb] 
\centering {
\includegraphics[width=0.45\textwidth]{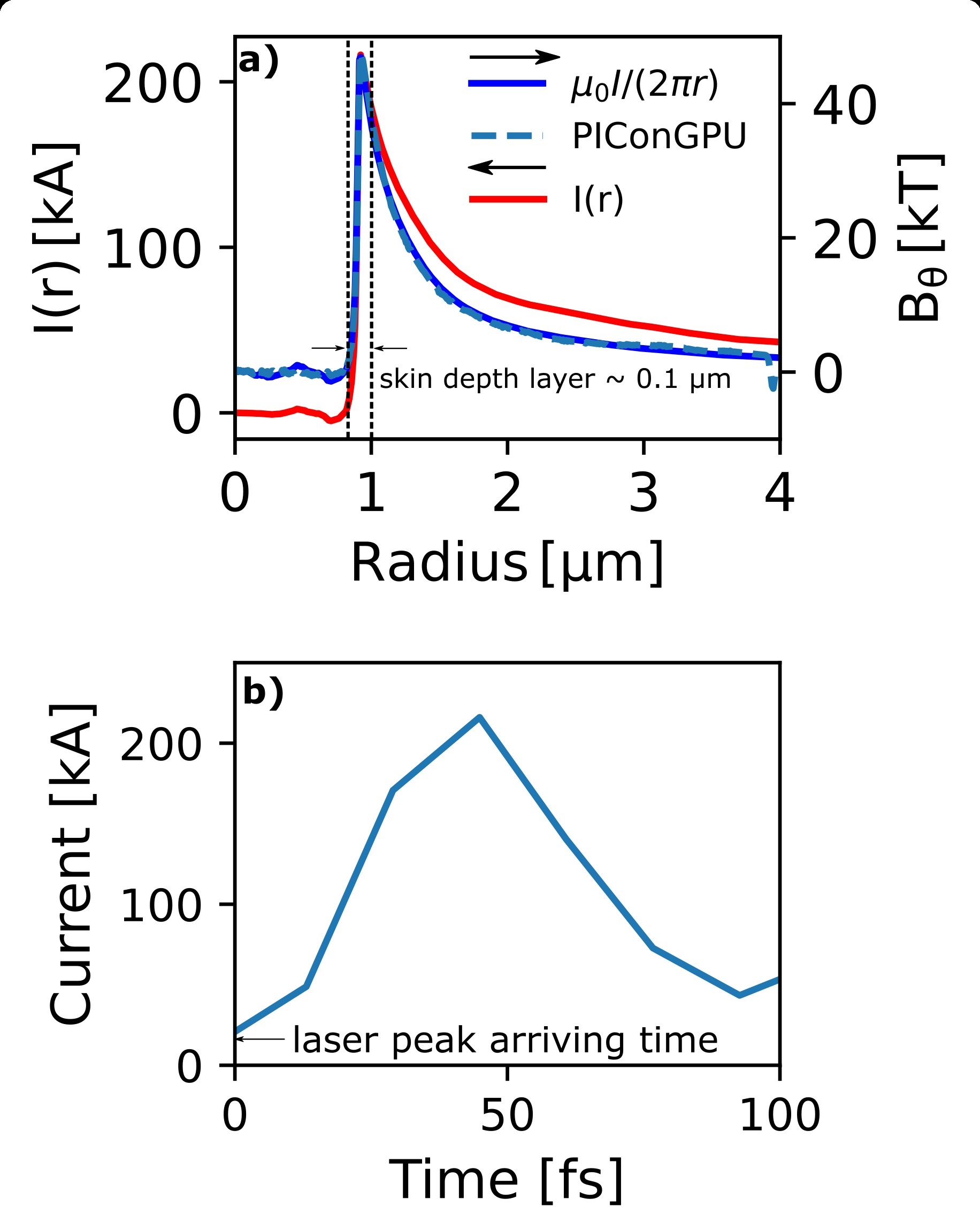}} 
\caption{a) Total current and magnetic field distribution in the radial direction at y = \SI{5}{\um} in PIConGPU simulation at 45 fs. b)Surface return current component evolves with time at an offset of \SI{5}{\um} to the laser spot in 3D PIConGPU simulations.}
\label{Fig.surface_current_magnetic}
\end{figure}

\begin{figure*}[htb] 
\centering {
\includegraphics[width=0.9\textwidth]{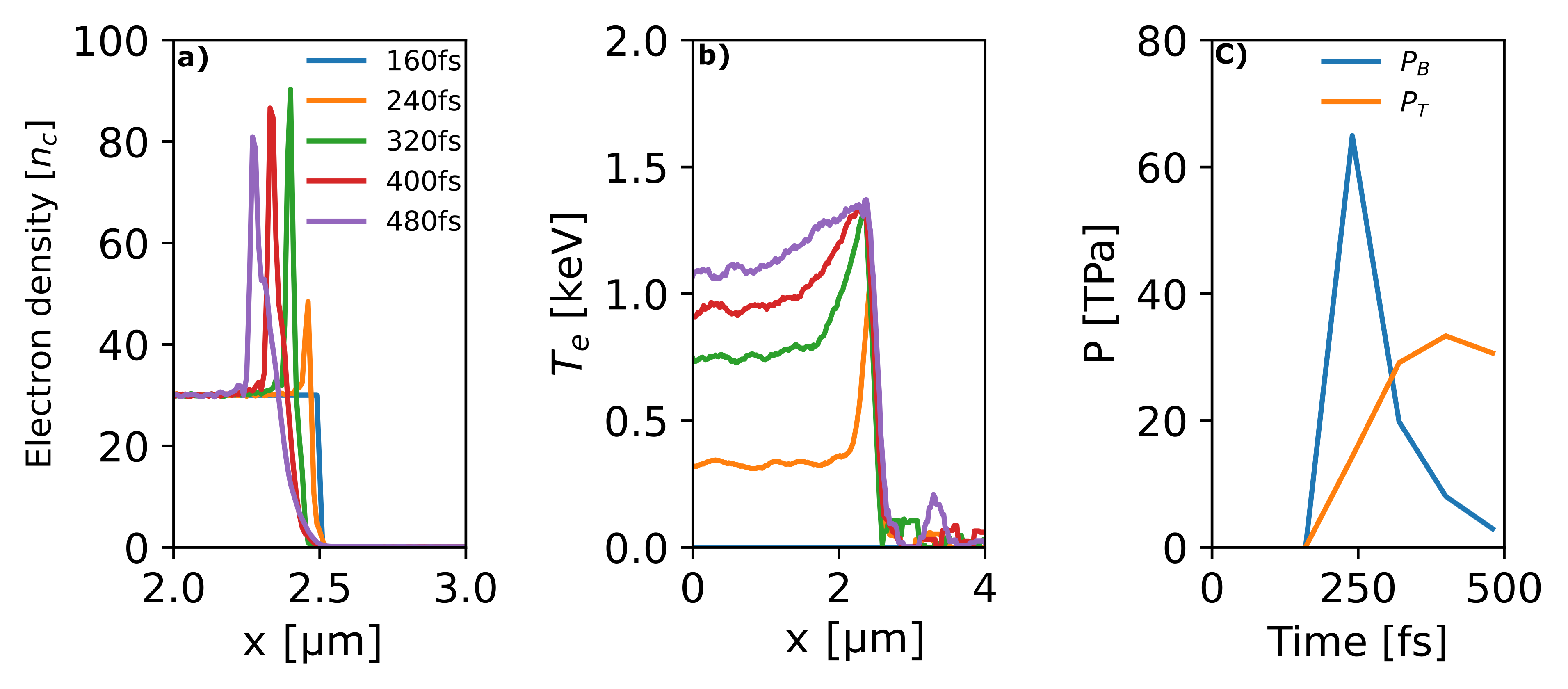}} 
\caption{Lineout of Physics quantities at y=\SI{60}{\um} in the 2D PIC simulation. The geometry of the simulation box is shown in figure \ref{Fig.picls_jy10}. The x=\SI{0}{\um} indicates the center of the hydrogen slab as shown in figure \ref{Fig.picls_jy10} (a) - (b). a) The hydrogen electron density distribution at different times. b)Electron temperature distribution at different times. c) The magnetic pressure and plasma thermal pressure varies with time on the plasma surface.}
\label{Fig.PICLS_2D_compression}
\end{figure*}

Figure \ref{Fig.surface_current_magnetic}(a)  shows the current at an offset of \SI{5}{\um} to the laser spot from the 3D PIC simulations (see Appendix A).
The backward surface return current $I_{re}$, which is distributed in the \SI{0.1}{\um} skin depth, increases dramatically to 210 kA at the plasma surface. The decrease of the integrated current density is observed when the radius exceeds \SI{1}{\um}, indicating the existence of the forward laser-accelerated electron current $I_{h}$. The current drops to 0 when the radius is \SI{4}{\um}, showing that the $I_{h}$ and the $I_{re}$ are approximately equal to each other. The obtained backward surface return current $I_{re}$ varies with time at y = \SI{5}{\um} are shown in figure \ref{Fig.surface_current_magnetic}(b). A backward surface return current with a duration of 100 fs is observed.
\\
The magnetic field generated by the current can be deduced using Biot–Savart's law 
\begin{equation}
B_{\theta}(r)=\mu_0I(r)/(2\pi r),    \, 
\end{equation}
\begin{equation}
    I(r)=\int_{0}^{r}2 \pi r j_y \,dr ,
\end{equation}
where r is the distance to the target center, and I(r) is the radial current distribution. $B_{\theta}(r)$ is compared with the magnetic distribution simulated by the 3D PIC simulation. $B_{\theta}(r)$ and magnetic field in 3D PIC simulation are marked as the solid and dashed curves, respectively, in figure \ref{Fig.surface_current_magnetic}. The maximum difference between the two magnetic fields is within \SI{1}{\%}, demonstrating that the surface magnetic field (r $\sim$ \SI{1}{\um}) is generated by the surface return current ($I_{re}$) and drives the z-pinch effect on the hydrogen plasma. Consequently, the surface return current can be utilized to analyze plasma surface heating and compression. \\
In the 2D PIC simulations, the region offset by \SI{60}{\um} from the laser spot is specifically selected to investigate the pure surface return current driven magnetic compression to the target. As exhibited in Figure \ref{Fig.PICLS_2D_compression}(a) and (b), lineout of electron density and electron temperature evolution in this area are presented. Here the bulk and surface return return current both contribute to the electron temperature \cite{yang2023,huang2022dynamics}. The surface return current heats the surface electron to 1.0 keV within the skin depth at 240 fs, after that, The rate of increase in surface electron temperature rises slowly due to the rapid decay of the surface return current after 240 fs.
The bulk return current heats the bulk electron temperature from 350 eV at 240 fs and to 1.1 keV at 480 fs.
The magnetic field pressure and thermal pressure are shown in figure \ref{Fig.PICLS_2D_compression}(c). Before 300 fs, the plasma $\beta = P_T/P_B \leq 1$, hence the transient magnetic pressure drives the density compression, up to a compression factor of 3.
This mechanism is consistent with the traditional Z-pinch effect. The $\textbf{J}\times \textbf{B}$ force driver only lasted for 100 fs, and the following shock wave velocity can be assumed to be constant in a short time. Based on the compression wavefront at respective times of 240 fs, 320 fs, 400 fs, and 480 fs, we calculate the inward velocity of the compression wave to be \SI{800}{\km/s}. The sound speed of the hydrogen is
\begin{equation}
 C_s=\sqrt{\gamma Zk_bT_e/m_i} \approx \SI{100}{[km/s]}
\end{equation}
scale where $\gamma=1.5$ is the adiabatic index. The magnetic field induced by the surface return current generates an initial inward shock wave with a Mach number of 8. Magnetic compression ceases beyond 300 fs, as indicated by a plasma $\beta \geq 1$ as shown in Figure \ref{Fig.PICLS_2D_compression}(c), demonstrating that the Z-pinch effect only dominates for the duration of the current pulse.\\
In this specific region of the showcase, bulk return current heating reduces the pressure gradient between the plasma surface and the bulk, so that the generation of the shock due to the surface return current heating is not visible. We aim to isolate a region where the plasma heating is solely attributed to the surface return current. However, it requires to simulation of a target with a transverse dimension extending to several hundred micrometers which is extremely challenging for the PIC simulations due to the highly demanded computational time. Consequently, to accurately estimate the compression effects of surface return current on the target, a method beyond the scope of PIC simulations is necessary.
\subsection{Heating mechanism of surface return current}
\begin{figure}[htb] 
\centering {
\includegraphics[width=0.45\textwidth]{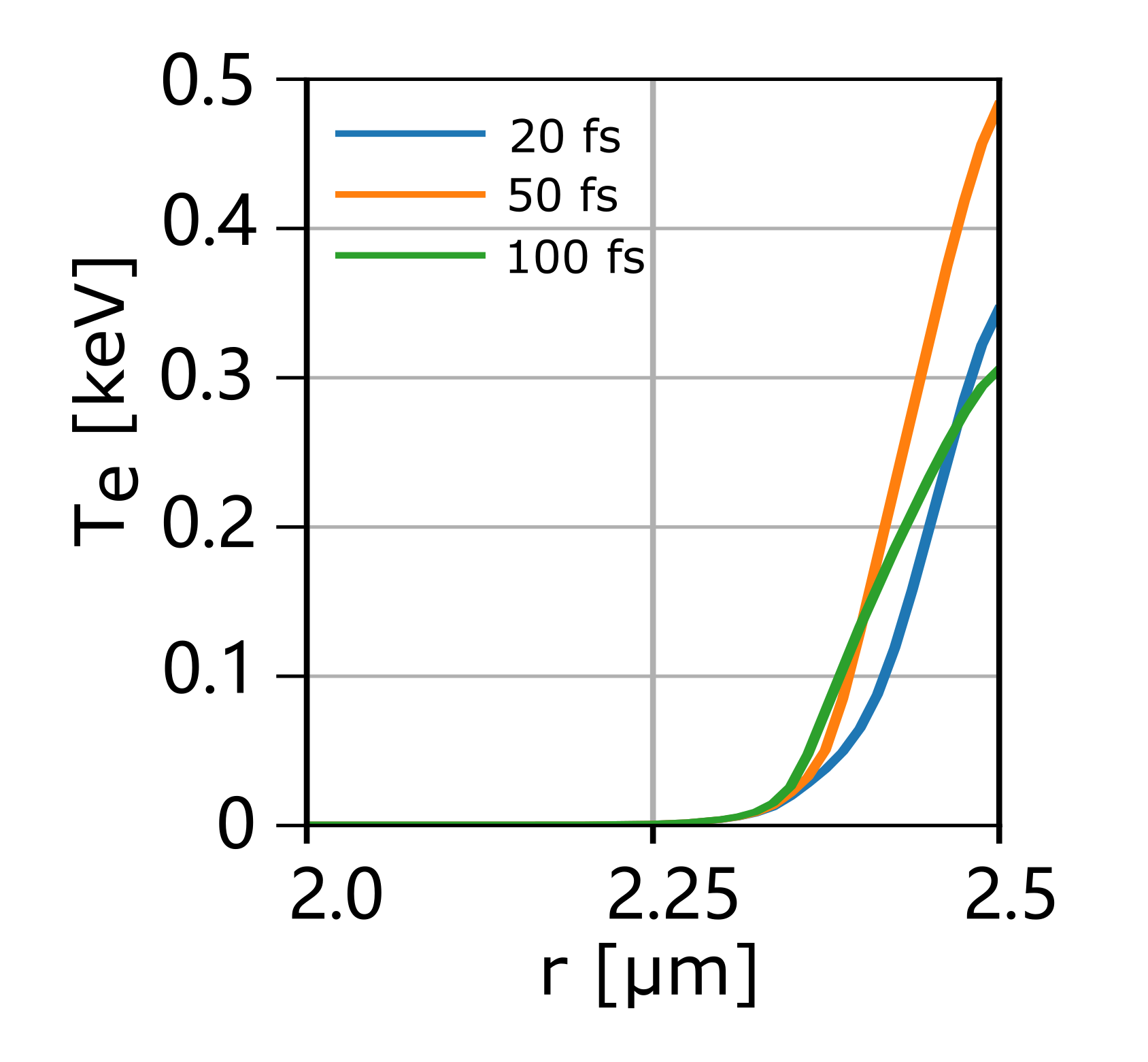}} 
\caption{The electron temperature distribution in the radial direction at times of 20 fs, 50 fs, and 100 fs.}
\label{Fig.return_current_heating}
\end{figure}
In this subsection, we explore the heating effects of the surface return current on a \SI{2.5}{\um} solid hydrogen target over a 100 fs timescale. The surface electron temperature influenced by this current can be deduced using the particle energy equation \cite{yang2023,kemp2006collisional}
 \begin{equation}
\frac{3}{2}n_c\frac{\partial T_e}{\partial t}=\frac{\partial}{\partial r}(K_{T_e}\frac{\partial T_e}{\partial r})+\frac{j_h(r)^2}{\sigma_{T_e}},\label{eq:heat_transfer}
\end{equation}
where $K_{T_e}$ is the thermal conductivity of cold electrons and $\sigma_{T_e}$ is the electron conductivity. $\tau_e$ is the bulk electron collision time. The equation's right side accounts for heat diffusion in the radial direction and the Joule heating due to the surface return current. 
The electron resistivity model and the heat diffusion coefficient are the function of temperature and density and are extracted from the SESAME equation of state \cite{johnson1994sesame}.

The distribution of current density in the radial direction is included in equation (\ref{eq:heat_transfer}) to study surface return current Joule heating in the plasma skin depth. Here the skin effect of the surface return current has to be considered. The skin depth under 100 eV electron temperature can be calculated as $\epsilon=\sqrt{2/\sigma_{T_e}\omega\mu_0} \approx \SI{0.1}{\um}$, where $\omega =2\pi/\tau $ is the frequency of the surface return current with $\tau$ being the duration time of the current pulse, and $\sigma_{T_e}$ is the resistivity of the hydrogen \cite{jordan1968electromagnetic}. Equation \ref{eq:heat_transfer} is solved in a 1D cylindrical geometry with an initial electron temperature of 1 eV, and the corresponding distribution of electron temperature at 20 fs, 50 fs, and 100 fs are depicted in Figure \ref{Fig.return_current_heating}. As the return current propagates through the target, joule heating of the surface plasma is observed. The peak temperature reaches 500 eV at \SI{50}{fs} and then decreases to 300 eV at \SI{100}{fs} caused by the heat diffusion within the \SI{0.1}{\um} skin depth layer.
Heat diffusion beyond the skin depth layer in the radial direction appears negligible in Figure \ref{Fig.return_current_heating}. As a result, bulk electrons remain relatively cool at \SI{100}{fs}. This differs from the PIC simulation in Figure \ref{Fig.PICLS_2D_compression} (b) where the bulk heating of the electrons is observed.  The plasma with high surface temperature is expected to generate several tens of terapascal pressure, which can also contribute to the formation of
converging shocks. This will be discussed in the subsection C.

\subsection{Dynamic compression in hydrogen jets initiated by the z-pinch and surface ablation}
In this subsection, we combine the effects of both z-pinch and surface heating to explore convergent shock compression over 100 ps time scales. However, the magnetic pressure is still observed after 480 fs as shown in Figure \ref{Fig.PICLS_2D_compression} (c), showing the existence of a magnetic field. To estimate the influence of the remaining magnetic field on the convergent shock evolution in ps time scales, we utilize the Magnetic Reynolds number which is an indication of the magnetic diffusion effect and magnetic frozen effect. 
The Magnetic Reynolds number, which gives an estimate of the effects of magnetic advection to magnetic diffusion, is written as $R_m = UL/\eta \approx 9.2$, where $U \approx \SI{300}{km/s}$ is the characteristic velocity, $L = \SI{5} {\um} $ is the characteristic length, and $\eta=(\mu _0 \sigma_{T_e})^{-1}$ is the magnetic diffusivity. The Magnetic Reynolds number $R_m > 1$ indicates that the magnetic field will be frozen in the ablation plasma and will have no influence on the shock wave evolution in the bulk plasma. Therefore, the magnetic diffusion effect is excluded from this study. Consequently, a hydrodynamic method can be applied. \\

We aim to investigate the pure surface return current compression without bulk return current heating on the hydrogen. 
Assuming the bulk plasma in infinite far away from the laser spot is cold with an initial temperature of 1 eV and there is no decay for the surface return current density when the surface return current pulse propagates along the wire. The z-pinch and surface heating effects provide the initial kinetic energy and thermal energy of the hydrogen plasma, respectively. Therefore, the electron temperature in figure \ref{Fig.return_current_heating} and shockwave velocity are the initial conditions of hydrodynamic simulations. 
The radiation cooling of hydrogen at a 300 eV scale is \SI{0.14}{eV} per electron and is excluded from our considerations \cite{yang2023}. We solve 1D cylindrical, one-fluid, two-species (ion and electron), and two-temperature hydrodynamic equations using the FLASH code \cite{fryxell2000flash,dubey2009extensible}. The hydrogen SESAME equation of state \cite{johnson1994sesame} is applied. \\

\begin{figure}[htb] 
\centering {
\includegraphics[width=0.5\textwidth]{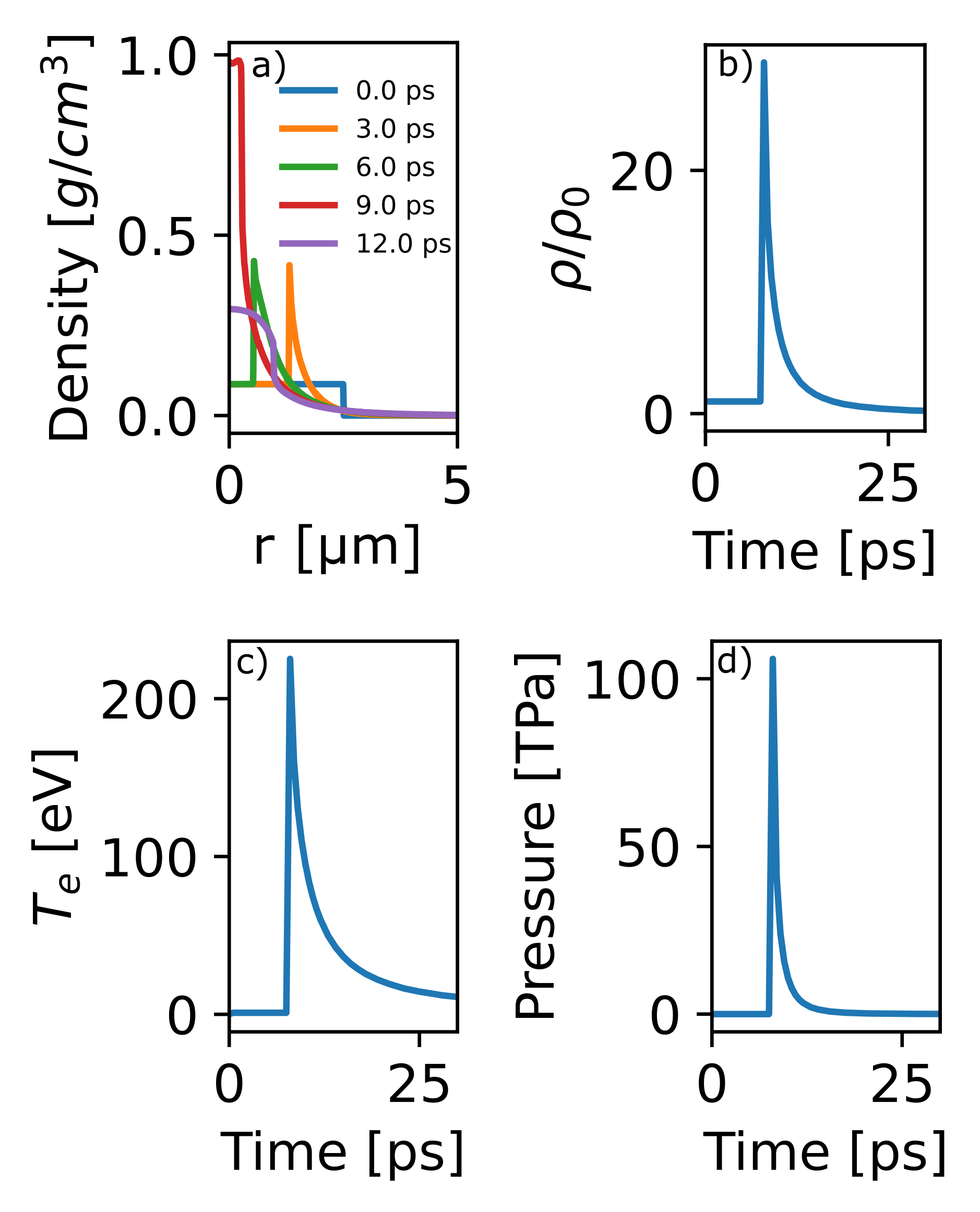}} 
\caption{a) The hydrogen density distribution along the radius at different times. b)-d) The density compression factor, electron temperature, and pressure of hydrogen vary with time at the target center respectively.}
\label{Fig.hydrogen_a20_draco_20230914}
\end{figure}

The results of the simulations are shown in Figure \ref{Fig.hydrogen_a20_draco_20230914} (a). It is evident that converging shock compression has three distinct phases. The initial phase features the expansion of the hot ablation plasma into the surrounding space, along with the formation of a converging shock wave. This shock wave moves inward at an average velocity of 700 km/s during the initial 2 ps. The density compression factor in this period ranges from 4 to 5. As the shock wave moves inward, more mass is accelerated by the shock wave. A decrease in the ablation plasma's temperature to 50 eV due to the expansion around 5 ps is also observed. Both factors contribute to the reduction of shock velocity to roughly 400 km/s. The subsequent phase witnesses the converging shocks reaching the target's center and collapsing around 8.5 ps. The density compression factor can reach a factor of 29 and the peak pressure can reach 100 TPa with 220 eV temperatures in a plasma core as shown in Figure \ref{Fig.hydrogen_a20_draco_20230914} (b) - (d). The final phase is characterized by the converging shock waves rebounding after reaching maximum compression, and subsequently propagating outwards. At around 40 ps, the reflected shock waves reach the interface of the plasma expansion, enhancing the expansion speed from 100 km/s to approximately 250 km/s. The corresponding plasma temperature is decreased to a few eV.

The simulation results show that the shock wave is initiated by the magnetic field compression and ablation plasma heated by the Joule heating associated with the return current. In this showcase, magnetic field compression (plasma $\beta < 1$) is the dominant mechanism to initiate the shock wave. 
Due to the short duration time of the return current, the energy can only deposit in the skin depth of the target, resulting in a clean and symmetry converged shock compression to the target to extremely high density and pressure. In our following section, we extrapolate our theory to targets with higher atomic numbers (Z) and larger radii to explore the broader applications of dynamic compression steered by the return current methodology.

\subsection{Scaling law of return current compression to high Z and large radius solid target}
In this subsection, we extend the surface return current density to arbitrary atomic number Z and target radii to investigate the initiation mechanisms of convergent shocks via surface return current.
The 2D and 3D PIC simulations show that the surface return current is quite similar with different target radii. Hence, we propose that the surface return current is mainly governed by the incident laser, with the return current duration and the corresponding skin depth remaining constant for fixed laser parameters.
Based on the equation (2),
for different radius targets, the scaling of the surface return current density can be derived by
\begin{equation}
    \frac{j_1}{j_2} \approx \frac{r_2}{r_1}.
\label{J_r}
\end{equation}\\
The corresponding magnetic field  on the target surface can be derived by
\begin{equation}
    \frac{B_1}{B_2} \approx \frac{r_2}{r_1}.
\label{B_r}
\end{equation}\\
From the 2D and 3D simulation results those use the wire radius of 2.5 $\mu$ m and 1 $\mu$ m respectively, it can be seen that the ratio of the maximum magnetic field on the plasma surface is $ B_{2max}/B_{1max}= \SI{18}{kT}/\SI{46}{kT} \sim r_{1}/r_{2} = 0.4 $, which shows that the 2D and 3D PIC simulations are consistent with each other.\\
A scaling law can be derived for various materials with different radii based on equations (\ref{eq:heat_transfer}), (\ref{J_r}), and (\ref{B_r}). The heat transfer term in equation (\ref{eq:heat_transfer}) is neglected, and the surface return current of \SI{2.5}{\um} is taken as the input. The transient peak electron temperature on the plasma surface with a 30fs, \SI{5e20}{W/cm^2} laser is approximately given by
\begin{equation}
    T_e=2.25(\frac{\SI{30}{n_c^{800nm}}}{n_i})^{0.4}(\frac{\SI{2.5}{\um}}{r})^{0.8} \SI{}{[keV]}.
\end{equation}\\
The peak magnetic pressure, which varies with the target radius, is expressed as
\begin{equation}
    P_B=101.6(\frac{\SI{2.5}{\um}}{r})^2 \SI{}{[TPa]}.
\end{equation}\\
In the dynamics of a surface return current interacting with a wire target, the peak values of $P_B$ and $P_T$ aren't reached simultaneously. Nevertheless, given the transient character of the surface return current, the plasma $\beta$ remains a reliable metric to discern which mechanism predominantly initiates a convergent shock within the wire target.
\begin{equation}
    \beta=\frac{P_T}{P_B}=\frac{Zn_ik_BT_e}{P_B}=\frac{Zn_i^{0.6}}{41.6}(\frac{\SI{2.5}{\um}}{r})^{-1.2},
\end{equation}\\
\begin{figure}[htb] 
\centering {
\includegraphics[width=0.45\textwidth]{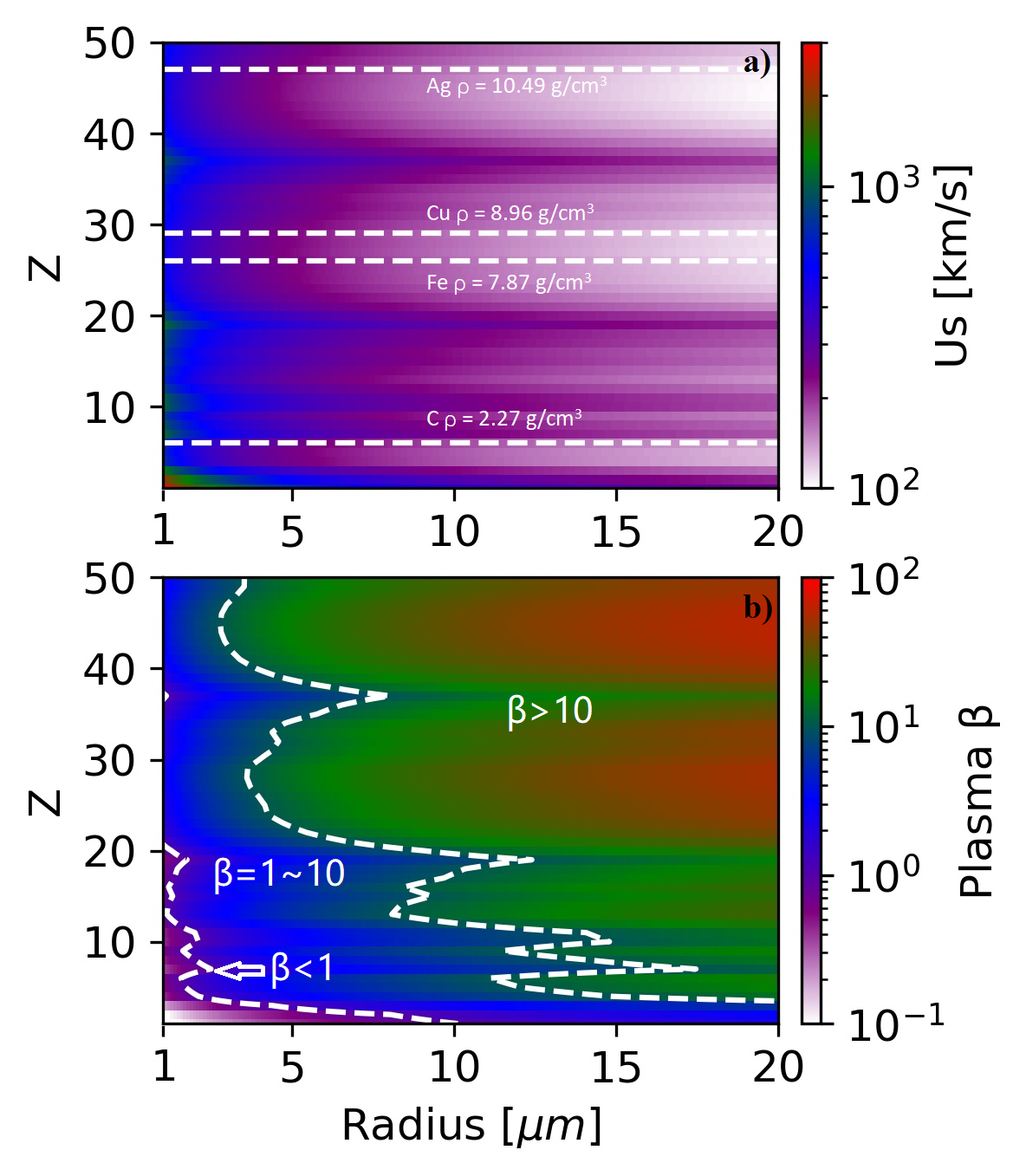}} 
\caption{a)The transient shock velocity map varies with the target radius and atom number Z. b) The plasma $\beta$ varies with the target radius and atomic number Z. The two contour curves indicate the $\beta$ = 1 and $\beta$ = 10.}
\label{Fig.scaling}
\end{figure}
where $n_i$ is the target atom density with the unit of critical density corresponding to the laser wavelength of \SI{800}{nm}, and Z is the target atomic number. It can be observed that the plasma $\beta$ is proportional to the target radius times the atomic number with a factor of - 1.2 and 0.6 respectively. The plasma $\beta$ map as a function of Z and radius is shown in Figure \ref{Fig.scaling}(b). Two contour curves indicate $\beta$ = 1 and $\beta$ = 10. Here we use the copper wire targets with the radii of \SI{10}{\um}, \SI{20}{\um}, and \SI{30}{\um} as representative examples. The corresponding plasma $\beta$ values of 38, 87, and 142 indicate that thermal pressure, induced by the surface return current joule heating, dominates the surface plasma pressure, and initiates a convergent shock.
Taking the surface return current scaling law, we can estimate the peak surface return current density as \SI{70}{kA/\um^2}, \SI{35}{kA/\um^2}, and \SI{23.3}{kA/\um^2} respectively. By using the equation \ref{eq:heat_transfer}, the peak surface temperature is calculated as 250 eV, 100 eV, and 50 eV respectively. Using the hydro simulation code FLASH with copper SESAME equation of state \cite{johnson1994sesame}, the peak pressure at the target center compressed by the surface return current is 125 TPa, 5.3 TPa, and 1 TPa respectively. The figure \ref{Fig.PIC_copper} (a) shows the copper density map varied with time and radius for a \SI{10}{\um} radius copper wire. A convergent shock compression is observed. The convergent shock is imploded at about 550 ps. The peak pressure is 125 TPa as shown in the figure \ref{Fig.PIC_copper} (b). Figure \ref{Fig.PIC_copper} (c) - (d) shows the electron energy density and the surface return current density in a \SI{10}{\um} width copper slab irradiated by a $a_0 = 15$, 30 fs and \SI{3}{\um} laser (The details setup of the PIC simulation can be found in Appendix A). The surface return current heating happens in about \SI{0.1}{\um} skin depth. The compression of the ion density is not observed in the first 100 fs after the peak laser pulse. The simulation results show that the thermal pressure is about 10 times of the magnetic pressure at 107 fs.

With the ideal gas equation of state, the transient shock velocity can be derived based on the Rankine-Hugoniot relations:
\begin{equation}
    U_s=U_1+c_s\sqrt{1+\frac{\gamma+1}{2\gamma}(\frac{p_2}{p_1}-1)},
\end{equation}
where $c_s=\sqrt{\gamma p_1/\rho_1}$ is the upstream sound speed, and $p_1$ and $\rho_1$ are the pressure and density of the upstream with an electron temperature set to 1 eV (cold plasma). $p_2=P_B+P_T$ is the downstream pressure. $\gamma$ = 1.5 is the constant ratio of specific heat. $U_1$ is the upstream velocity and is set to zero due to a highly transient return current. The obtained shock velocity map varying with target radius and atomic number Z is shown in Figure \ref{Fig.scaling}(a).\\
Our findings reveal that, in low-Z and small-radius targets, magnetic pressure stands out as the principal compression mechanism. Yet, the interplay of the $\textbf{J}\times \textbf{B}$ force introduces Magnetohydrodynamic Rayleigh–Taylor (MRT) instabilities onto the plasma surface \cite{ong2023ultra}. Such instabilities disrupt the symmetry inherent to a convergent shock, thereby diminishing the attainable pressure—particularly in low-Z and small-radius configurations \cite{sinars2010measurements,weis2015coupling}. Utilizing the established framework, we deduced the MRT instability growth perturbation amplitude ratio as
\begin{equation}
  \eta/r=\frac{\eta_0}{r}e^{\int_{0}^{t} \sqrt{gk}\,dt} \propto \frac{1}{r}e^{r^{-3}\rho_0^{-1.5}},
\end{equation}
where $\eta_0$, $g$, and $k$ are initial instabilities amplitude, acceleration, and instabilities wave number, respectively. 
Based on the above analysis, we propose the use of High-Z and larger radius targets. In such configurations, ablation plasma significantly leads the initiation of a clean dynamic shock compression, while simultaneously, the occurrence of instabilities is reduced. The transient pressure can produce a shock with velocities ranging from \SI{100}{km/s} to \SI{1000}{km/s}, substantially larger than the sound speed, within an atomic Z number up to 50 and a radius up to \SI{20}{\um}. This evidence demonstrates that surface return current shock compression is a promising technique for generating transient high pressure in picoseconds time scales.
\begin{figure*}[htb] 
\centering {
\includegraphics[width=0.9\textwidth]{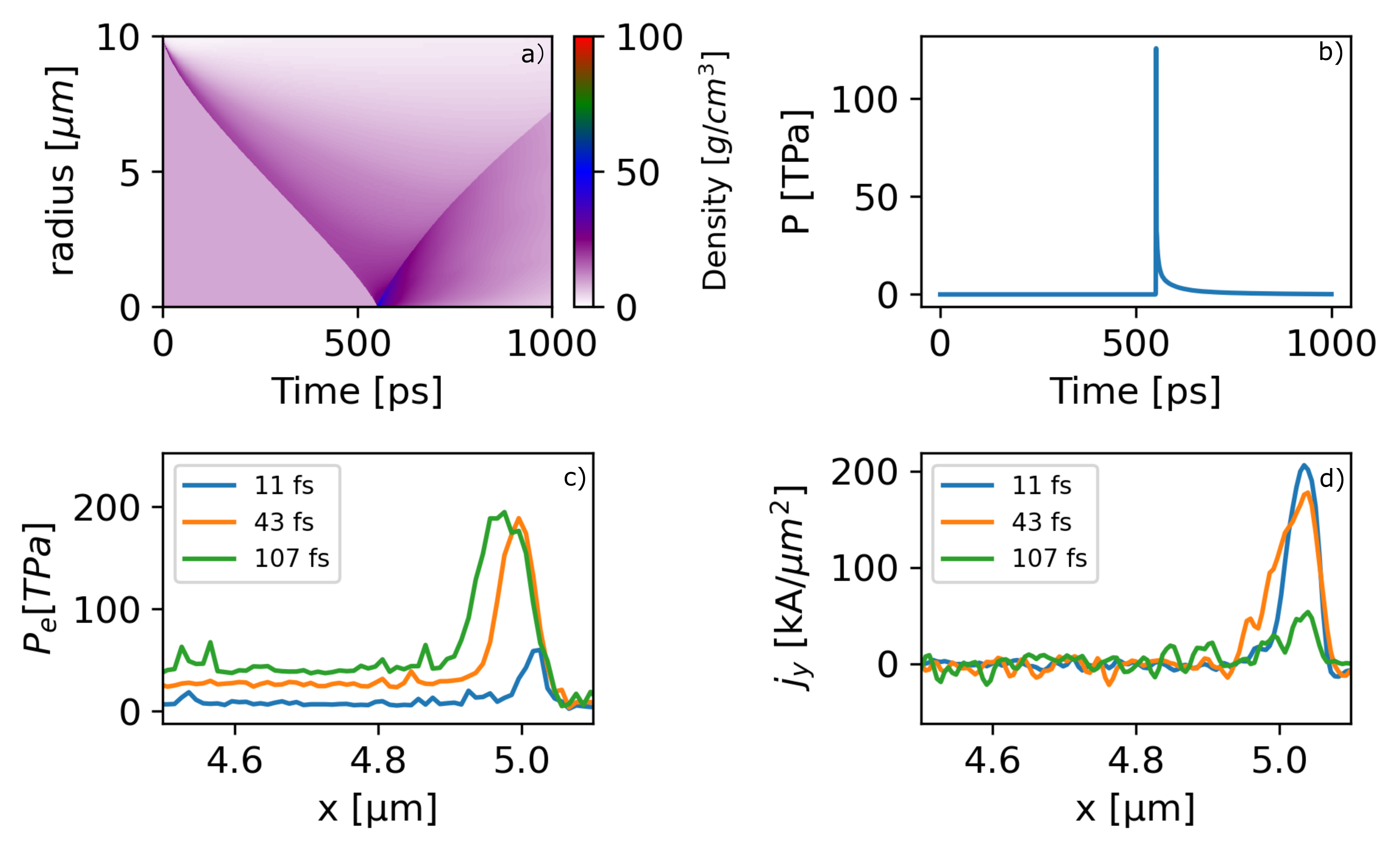}} 
\caption{a)The density variation over time and radius in a \SI{10}{\um} radii copper wire, as simulated by hydrodynamic simulation. b) The evolution of pressure versus time at the center of the copper target, as simulated by hydrodynamic simulation. c) The electron pressure calculated by $P_e=n_ek_BT_e$ over time in a \SI{10}{\um} width slab, as simulated by the PIC method. Here, x = \SI{0}{\um} represents the center position of the slab target. The offset is \SI{10}{\um} from the laser incident position. Time 0 marks the point when the maximum laser intensity reaches the target front surface. d) The corresponding evolution of surface return current density distribution at different times.}
\label{Fig.PIC_copper}
\end{figure*}

\section{Indications from experiments}
\begin{figure*}[htb] 
\centering {
\includegraphics[width=0.9\textwidth]{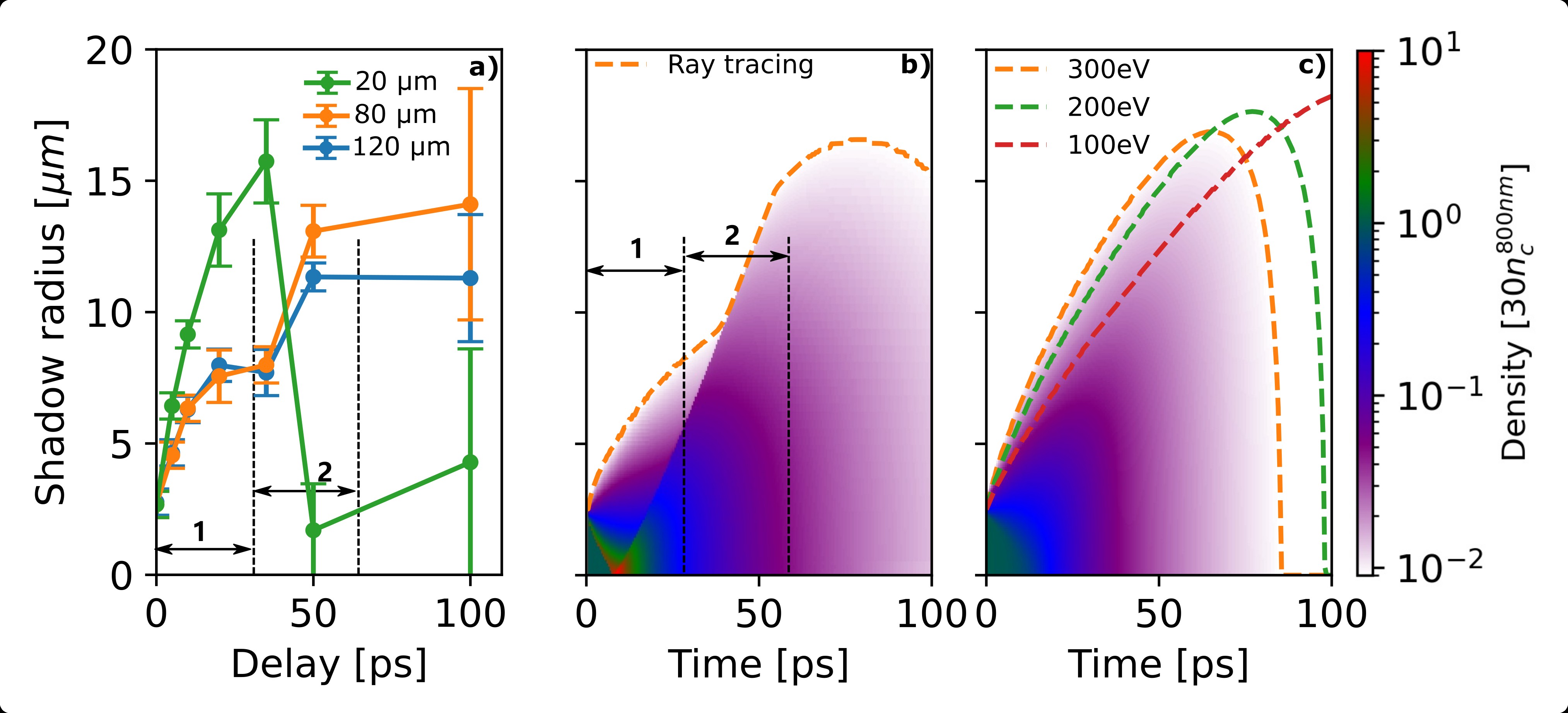}} 
\caption{a) Experimental shadowgraphy radius of the target. The labels with different lengths indicate the distance to the laser spot. The uncertainty, as represented by the error bar, arises from 1. the statistical variations across effective multiple experimental runs. Run numbers for each delay (from 0 ps, 5 ps, 10 ps, 20 ps, 35 ps, 50 ps to 100 ps) are 1, 1, 2, 3, 4, 2, and 2, respectively. 2. due to the resolution limit (\SI{2}{\um}) of the probing. 3. due to different laser target overlaps, not all shots have the same initial conditions.
b) The map shows the hydrogen density distribution at different times simulated by the hydrodynamic method in section II C. The dashed line shows the shadow diameter obtained by the ray tracing method. c) The color map shows the density evolution with time in a uniform isochoric heating process with an initial temperature of 300eV. The dashed lines show the ray tracing results with different initial temperature conditions.}
\label{Fig.shadow_diameter_a20}
\end{figure*}
As discussed in Section II, the converging shock compression initiated by the surface return current is characterized by three significant effects:  hot surface plasma, cold bulk plasma, and the reflected shock wave. Each of these effects plays an essential role in understanding and analyzing the phenomena associated with convergent shock compression.
These effects can be reflected in the plasma density distribution, which can be probed by the optical laser shadowgraphy technical \cite{yang2023, Martin2022, Bernert2022}. Experiments conducted at the HZDR 150TW Draco laser facility previously published in \cite{Martin2022} could provide valuable insights into the occurrence of converging shock compression in hydrogen jets. In these experiments, a solid hydrogen jet with a radius of \SI{2.5}{\um} is irradiated by a \SI{5e20}{Wcm^{-2}} intensity, 30 fs short pulse laser. Figure \ref{Fig.shadow_diameter_a20} (a) presents the shadowgraphy diameter of the solid hydrogen taken from Figure 5.3 (row 3) in reference \cite{Martin2022} at various time delays relative to the pump laser. The different colored curves represent the shadow diameters at different distances from the laser spot. Here, we use the offset of \SI{120}{\um} as a reference to the simulations. The bulk temperature at \SI{120}{\um} is very similar to the initial condition of simulation at Section II C (see details in Appendix B). Ray tracing simulations bridge the experimental data and computational predictions \cite{yang2023, Bernert2022}, indicating the surface return current's role in initiating converging shock compression on hydrogen jets.

The ablation plasma: as shown in region 1 in Figure \ref{Fig.shadow_diameter_a20} (a) and (b), a rapid shadow diameter expansion rate within the initial 20 ps indicates the presence of a hot corona plasma. Assuming a constant critical density for a 515 nm laser probe, the hot ablation plasma is expected to expand at an ion sound speed of $v=\sqrt{\gamma k_BT_e/m_i} \propto T_e^{0.5}$. The expansion velocity can be estimated from the initial slope of the shadow diameter. We use the shadow diameter \SI{120}{\um} offset to the laser spot, where the bulk heating is small. The slope can be calculated as 360 km/s. With a calculated temperature of 300 eV (see figure \ref{eq:heat_transfer}), the corresponding expansion velocity is 207 km/s, which is roughly consistent with the experimentally observed slope. \\

The plasma transparent time: the transparency of the plasma in the shadowgraphy image is not observed. This is distinct from the isochoric heating process\cite{yang2023}. In the ray tracing simulations, we can identify the apparent difference between these two conditions. Figure  \ref{Fig.shadow_diameter_a20} (c) displays the shadow diameter of adiabatically expanded plasma with temperatures of 100 eV, 200 eV, and 300 eV, respectively. The expansion shadow diameters of isochoric heating with electron temperatures 200 eV and 300 eV cases have a good agreement with the experimental data in the first 20 ps. However, both shadow diameters became transparent at 80-100 ps, indicating that the bulk plasma temperature is overestimated with the 200-300 eV isochoric heating assumption. When we decrease the initial temperature to 100 eV, the transparency of the plasma is not observed, but the shadow diameters in the first 20 ps are under the experimental data, showing the 100 eV temperature is underestimated. The comparison suggests that the bulk plasma temperature is significantly lower than that of the corona plasma. This is consistent with the surface return current heating mechanism described in Section II. 

The reflected shock wave: as shown in figure  \ref{Fig.shadow_diameter_a20} (a) and (b) region 2, the shadow diameter increased dramatically. Since the corona plasma density decreases due to expansion, the shadow diameter expansion rate also decreases. This can be observed from \SI{20}{ps} to \SI{40}{ps}. A smooth shadowgraphy diameter curve should be obtained \cite{yang2023}, indicating a diabatic expansion process. However, the generation of a converged shock wave and rebounding of the shockwave can introduce high-density plasma into the expansion interface, increasing the corona plasma density and, consequently, the shadow diameter expansion rate increase again. Therefore, the second increase in shadow diameter is a sign of a rebounded shockwave.\\ 

In summation, the experimental shadowgraphy images vividly portray the hallmarks of dynamic converging compression in hydrogen jets, findings corroborated by ray tracing simulations. For a more comprehensive understanding, refining the experimental data is paramount—specifically, minimizing uncertainties, enhancing temporal resolution at around 40 ps to chart the reflected shock's evolution, and diversifying radii investigations using metal wires. Leveraging X-ray free electron lasers could furnish direct insights into the shock wave. Notably, our analysis presupposes a 1D symmetric compression, a simplification that may not encapsulate the real process's intricacies. As such, perfect congruence between experimental data and simulations remains elusive.

\section{Outlook and Conclusion}
In summary, we have demonstrated a novel method for inducing converging shock compression by utilizing transient surface return current during femtosecond short pulse laser-target interactions. 
The surface return current is studied by 2D and 3D PIC simulations. With a \SI{5e20}{W/cm^2}, 30 fs short pulse incident laser, the surface return current shows a duration of approximately \SI{100}{fs}, reaching a maximum current of \SI{210}{kA}. 
Our analytical approach examines the impact of the return current on wire targets, focusing on the effects due to the $\textbf{J}\times \textbf{B}$ force compression and Joule heating. Intriguingly, the mechanism of convergent shock compression finds itself heavily influenced by the plasma $\beta$. Here, magnetic field compression takes the lead when plasma $\beta < 1$.
Considering the case of a \SI{2.5}{\um} hydrogen wire: the current density confines itself within a skin depth of roughly \SI{0.1}{\um}, displaying a peak current density of \SI{0.28}{MA/\um^2}. The hydrogen density is compressed by a factor of 29, resulting in a peak pressure of 100 TPa and a temperature of 220 eV when plasma is imploded in the center. 
A scaling law is proposed to extend the implications of the return current to targets of varying atomic number (Z) and radius. The results suggest that, as the target radius and atomic number (Z) increases, surface ablation becomes the dominant mechanism to initiate the convergent shock wave. In the Z and radius ranges investigated ($Z\leq50$ and $r\leq20$ \SI{}{\um}), a shock wave with a velocity $U_s > $ 100 km/s can be initiated in all cases.
Utilizing PIC and hydrodynamic simulations, alongside experiments with \SI{2.5}{\um} hydrogen and optical shadowgraphy ray tracing simulations, we observed the rapid expansion of hot surface plasma, the presence of cold bulk plasma, and the reflective behavior of the shock wave. These findings indirectly indicate our hypothesis of convergent shocks initiated by the surface return current.
In summary, our discoveries shed light on the dynamics of return current-initiated converging shock compression during solid laser interactions. More than just a revelation, our work paves the way for innovative methods to realize high densities and pressures using joule-level short-pulse lasers.

\begin{acknowledgments}
FLASH  was developed in part by the DOE NNSA- and DOE Office of Science-supported Flash Center for Computational Science at the University of Chicago and the University of Rochester.\\

Author contribution statement: L.Y. built the theory. L.Y. and L.H. conducted the simulations. M.R., K.Z., and U.S. provided the experiment data. T.K., A.L., and T.T. analyzed the data. L.Y. wrote the publication. L.H. and T.C. supervised the project. All authors reviewed the manuscript. 
\end{acknowledgments}

\appendix
\section{PIC simulations}
This section introduces the 2D and 3D PIC simulations. 
Here, to investigate the surface return current generation process, we employ the PIC method using the 2D3V code PICLS \cite{sentoku2008numerical} and 3D3V code PIConGPU \cite{bussmann2013radiative}. A relativistic binary collision operator \cite{sentoku2008numerical,perez2012improved} is included to calculate electron and ion temperatures. The 2D and 3D PIC simulation will be used in Section II A to demonstrate the existence of the return current. The 3D PIC simulations will be used to quantitatively calculate the return current intensity and verify Ampere's law in a cylindrical coordinate system in Section II A. The 2D PIC simulations with longer simulation time, and larger simulation box will be used to show the dynamic process of surface return current compression as shown in Section II B.

For the hydrogen target, in the 2D PICLS simulations, the target is presumed fully ionized solid hydrogen with a density of $\SI{30}{n_c}^{800nm}$, where $\SI{}{n_c}^{800nm} = \SI{1.74e21}{cm^{-3}}$ is the plasma critical density corresponding to an 800 nm wavelength laser. The input laser is a Gaussian wave in time and space with a peak intensity of \SI{5e20}{W/cm^2} and a wavelength of \SI{800}{nm}. The FWHM duration of the laser is 30 fs and the FWHM spot size is \SI{3}{\um}. The laser propagates along the x-direction and is polarized in the y-direction. The simulation box size is \SI{15}{\um} in the x-direction and \SI{100}{\um} in the y-direction in terms of length, with corresponding grid numbers of 1500 $\times$ 10400. The target is placed between \SI{5}{\um} and \SI{10}{\um} in the x-direction to represent a \SI{5}{\um} diameter solid hydrogen jet. There is no scale length of the preplasma. The simulation time t=0   corresponds to the 
laser field on the target surface with peak intensity.
Field and particle-absorbing boundary conditions are applied to simulate an infinite-length hydrogen jet.

\begin{figure*}[htb] 
\centering {
\includegraphics[width=1\textwidth]{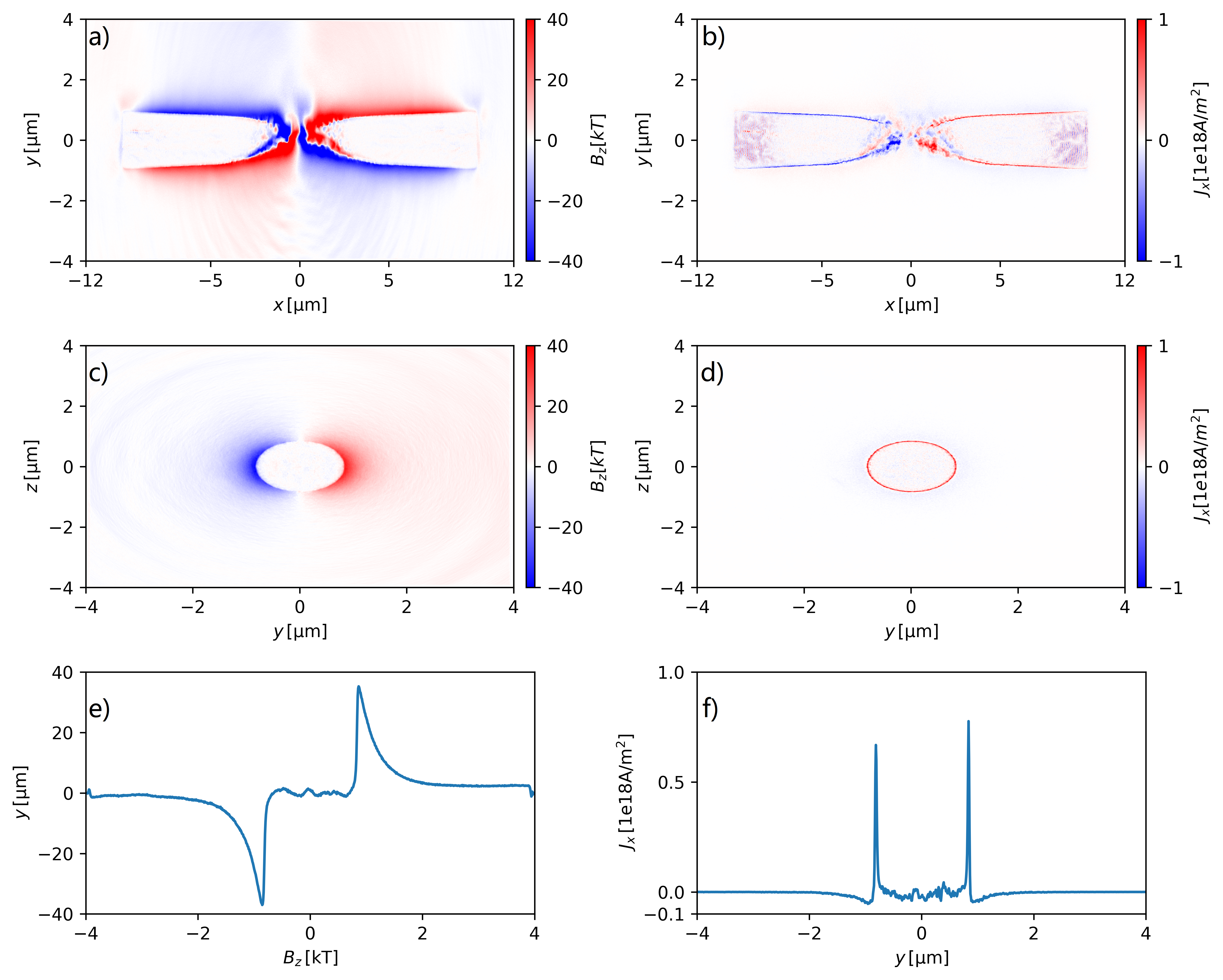}} 
\caption{PIConGPU 3D simulation of return current formation in a solid hydrogen wire at 100 fs. a) The 2D x-y distribution of magnetic field at z direction. b)The 2D x-y distribution of the current density in the x direction. c)The 2D y-z distribution of magnetic field at z direction. d)The 2D y-z distribution of the current density in the x direction. e) The lineout of $B_z$ at x=5 um in figure (a). f) The lineout of $j_x$ at y=5 um in figure (b).}
\label{Fig.3DPIConGPU}
\end{figure*}

The full 3D distribution of return current is simulated by PIConGPU. In the 3D simulations, the target radius is reduced to \SI{1}{\um}, and the length of the hydrogen jet is reduced to \SI{20}{\um} to decrease computation cost. The cell number is $960 \times 960 \times 2880$ in three directions. The remaining setup is identical to that of the 2D simulation. The results of 3D PIC simulations are shown in Appendix A.
The 3D PIConGPU simulation results are shown in Figure .\ref{Fig.3DPIConGPU}.

In the 2D PICLS simulations for the copper target, a copper slab is used with an initial electron and ion density of \SI{48}{n_c}. The slab dimensions are \SI{10}{\um} in the x-direction and \SI{26.7}{\um} in the y-direction with a simulation box of $\SI{40}{\um} \times \SI{26.7}{\um}$. The resolution of the cell is \SI{0.8}{\um}/150. The simulation includes the copper ionization process. The remaining setup parameters are identical to those used in the hydrogen simulations. The simulation results are shown in figure \ref{Fig.PIC_copper}(c) - (d).

\section{Bulk electron temperature along the longitudinal direction of the wire}

\begin{figure}[htb] 
\centering {
\includegraphics[width=0.45\textwidth]{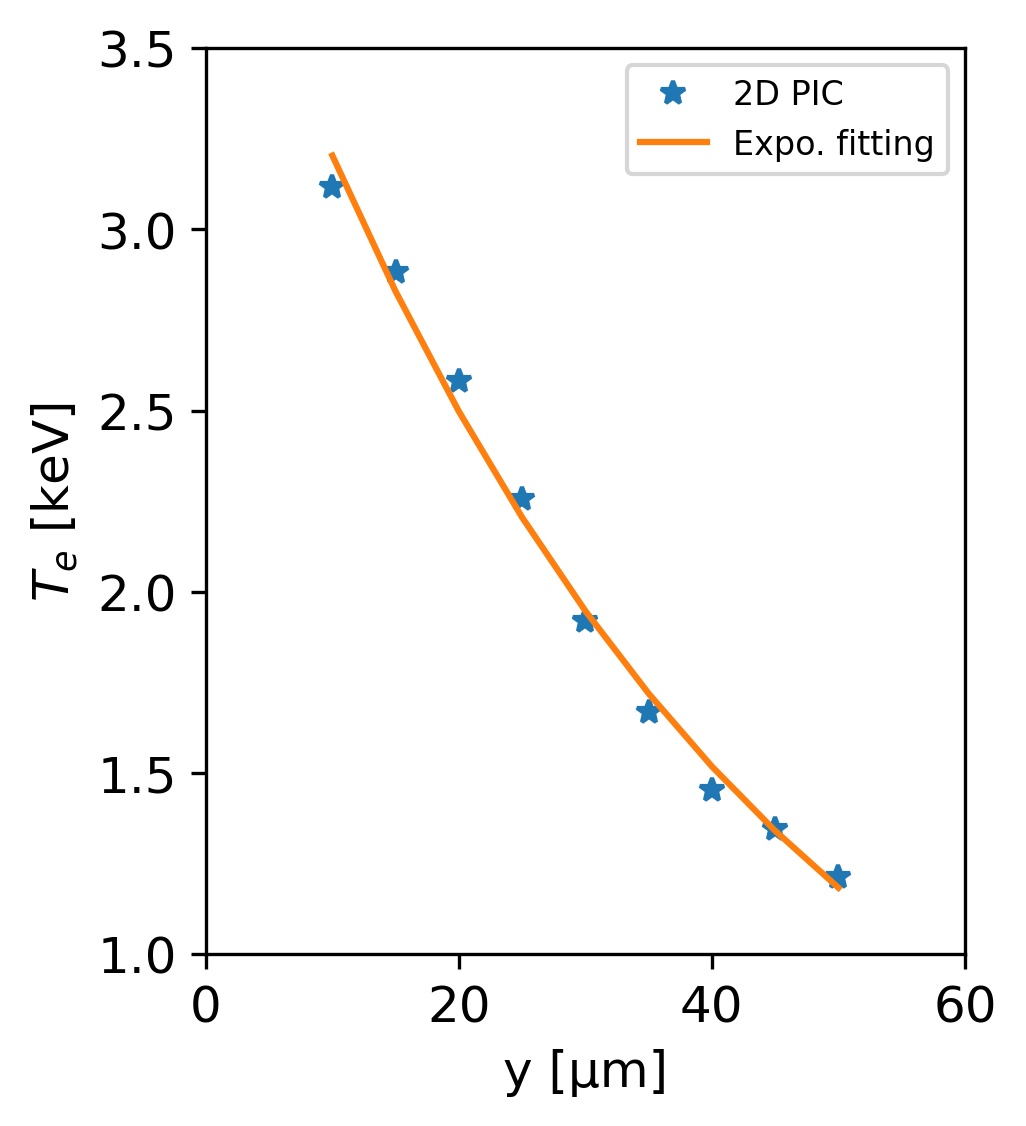}} 
\caption{The average bulk electron temperature distribution along the y-axis. The dot points are the results from 2D PIC simulations at 480 fs after the laser peak arrival time. The orange curve indicates the exponential fitting to the doted points.}
\label{Fig.PICLS_2D_tey}
\end{figure}

To estimate the bulk electron temperature \SI{120}{\um} to the laser-plasma interaction region, we take the 2D PIC simulations at 480 fs as the reference. Here, the electron temperature is derived as the $T_e = 2 E_z$, where $E_z$ is the electron average kinetic energy in the z direction \cite{yang2023}. The bulk electron temperature is taken as the $ 2 \times 2$ cell average temperature in the target center. The bulk electron temperature is shown in Figure \ref{Fig.PICLS_2D_tey}. It can be seen that the temperature distribution in the 2D PIC simulation follows an exponential decreasing function. Based on the fitted curve, the bulk electron temperature at \SI{120}{\um} offset can be estimated as 207 eV. Since the bulk electron temperature is overestimated in the PIC simulations \cite{yang2023}, we roughly estimated the overestimation factor of bulk electron temperature as 1 magnitude. Therefore, the bulk electron temperature is around 20 eV at an offset of \SI{120}{\um}. Compared to the strong Lorentz force and surface electron temperature caused by the surface return current, the thermal pressure of 20 eV can be neglected. As a simplification, the bulk electron temperature at \SI{120}{\um} is assumed as 1 eV.

\section{Experimental shadowgraphy radius evolution}
As additional information of figure \ref{Fig.shadow_diameter_a20} (a), here we take a lineout of every \SI{20}{\um} to plot the shadow radius.
\begin{figure}[htb] 
\centering {
\includegraphics[width=0.45\textwidth]{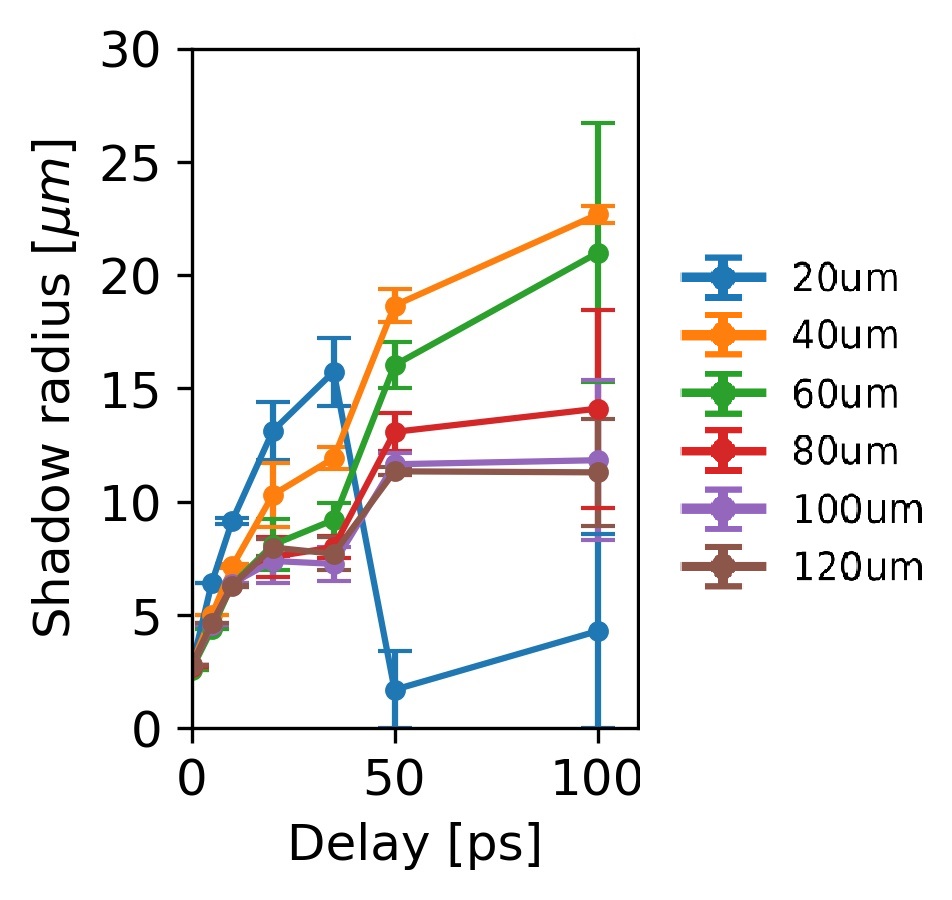}} 
\caption{Experimental shadowgraphy radius of the target. The labels with different lengths indicate the distance to the laser spot.}
\label{Fig.RT-33}
\end{figure}

\clearpage
\bibliography{ref}
\end{document}